\newcommand{\CLASSINPUTinnersidemargin}{16.9mm}
\newcommand{\CLASSINPUTtoptextmargin}{19mm}
\newcommand{\CLASSINPUTbottomtextmargin}{25mm}
\documentclass[conference,a4paper]{IEEEtran}

\usepackage{graphicx}
\usepackage{amsmath}
\usepackage{xcolor}
\usepackage[bookmarks, bookmarksopen=true, plainpages=false, pdfpagelabels, pdfpagelayout=SinglePage, breaklinks = true]{hyperref}
\newcommand\TI{\mathit{TI}}

\newcommand{\email}[1]{\ttfamily\href{mailto:#1}{#1}}

\begin{document}
\title{Modeling the dynamics and control of power systems with high share of renewable energies}

% affihorblockA{School of Electrical and\\Computer Engineering\\
\author{\IEEEauthorblockN{Sabine Auer and Tim Kittel}
	\IEEEauthorblockA{Potsdam Institute for Climate Impact Research (PIK)\\
		Member of the Leibniz Association\\
		%P.O. Box 60 12 03\\
		D-14412 Potsdam, Germany\\
		Email: \email{sabine.auer@pik-potsdam.de}}}

% after pdf' not found: using draft setting.

%Telephone: (800) 555--1212, Fax: (888) 555--1212}


\maketitle

\begin{abstract}
%\boldmath
A challenge for renewable and hybrid power systems is the dynamically stable integration of Renewable Energy Sources (RES). This paper specifically investigates the influence of intermittent RES and measurement delays from power electronic resources on frequency stability. In addition it presents an Open-Source framework to undertake dynamic RES modeling.

First, for local intermittent fluctuations in lossy distribution grids I find a remarkable and subtle but robust interplay of dynamical and topological properties, which is largely absent for lossless grids.

Second, I show how delays may induce resonance catastrophes and how the existence of critical delays sets an upper limit for measurement times. Further, I investigate whether centralized vs. decentralized power production, for different grid topologies, changes this behavior.

Third, the code used for producing the above results is in the process of being published as an open-source Software framework called PowerDynamics.jl being developed in the programming language Julia. It will cover the rich novel dynamics caused by the integration of RES with the implementation of Differential Algebraic (DAEs), Delayed Differential (DDEs) and Stochastic Differential Algebraic Equations (SDAEs).

Altogether, this paper investigates the stability of future power grids moving towards integrating more aspects of renewable energy dynamics and presents an adequate modeling framework for RES integration studies.
\end{abstract}


% \fi
\IEEEpeerreviewmaketitle

\section{Introduction}
The increasing share of Renewable Energy Sources (RES) poses a wide range of challenges for power grid stability. Today, in Germany 90\% of all installed power from RES lies in distribution grids \cite{verteilnetzstudie}. With a growing number of especially wind and solar power plants, new dynamics are introduced into the lower grid layers which need to be understood. This is the basis for a grid-stable integration of RES.

This paper splits in three parts. Section \ref{sec:intermittency} focuses on the influence of intermittent RES on power quality \cite{auer2017stability,auer2018stability}. For renewable energy fluctuations we study the stability of the synchronous state \cite{arenas2008synchronization,dorfler2014synchronization} and the ability of the system to keep frequency fluctuations small, also called power quality.

Section \ref{sec:delay} shows the influence of measurement reaction delays from inverter control schemes on the transient stability of power grids, quantified by the so-called basin stability.

In Section \ref{sec:os-modeling} we illustrate how throughout our research on renewable grid dynamics we encountered limitations for our analysis by the available open-source software packages in Python. 
%This especially refers to the modeling of Differential Algebraic Equations (DAEs), Stochastic Differential Algebraic Equations and Delayed (SDAEs) Differential Equations (DDEs). 
Transitioning to the novel programming language Julia eliminates these previous limitations with respect to numerical solvers and computational performance. This paper outlines the present limitations occurring from state-of-the-art open-source software for modeling power grid dynamics. As a next step and outlook for future modeling this paper introduces the new Julia Software framework called PowerDynamics.jl that is about to be published \cite{tkittelWIW2018}. 
In Section \ref{sec:discussion} we will summarize the results of the two exemplary research subjects and show how PowerDynamics.jl can be used for further research advancements.

\section{Intermittent Renewables and Power Quality}
\label{sec:intermittency}

In the following, we consider the model cases of an islanded power grid with intermittent fluctuations localized at a single node and focus on how the position of the perturbed node in the network influences its ability to corrupt the power grid's power quality.

\subsection{Case Study}
\subsubsection{Islanded Power Grid}
In order to understand the abillity of a distributed system to maintain synchrony, we present the case of an islanded power grid. Islanded power grids play a role for the decentral provision of energy, but also as part of a safety and stability strategy to localize faults by partitioning the grid into autonomous units. This case study does not relate to a real-world power grid but builds on a synthetic power network with realistic parameterization.

With high RES penetration our grid of interest is dominated by inverters. Network nodes are to be considered as effective nodes with a mix of at least one grid-forming inverter, a number of grid-feeding inverters and demand \cite{schiffer2016survey}. For reasons of grid stability, grid-forming inverters are assumed to be widely deployed but smaller in number compared to grid-feeding inverters. 
Inverters and their power electronics may be programmed as Virtual Synchronous Machines by using a smooth droop control. Following \cite{schiffer2016survey} this then leads to the same equations for the voltage angle $\phi$ and frequency $\omega$ as in the classical Synchronous Machine model \cite{Nishikawa2015}  in terms of the (virtual) inertia $H$, power infeed $P$, (virtual) damping $\alpha$, line susceptabilities, $Y=G+jB$, and voltage magnitudes $U$ \cite{schiffer2013synchronization}:
\begin{align}
\label{eq:dynmics}
\dot \phi_i &= \omega_i\;,\nonumber\\
DM_i \dot \omega_i &= P_i + \Delta P_k(t) - D_i \omega_i - \sum\limits_{j=1}^n P_{ij}\;,\nonumber\\
P_{ij} &= U_i|Y_{ij}|U_j \left[ \sin\left(\alpha_{ij}\right) + \sin\left( \phi_i - \phi_j - \alpha_{ij}\right)\right]\;.
\end{align}
%The virtual inertia and damping for the network model is given by the low-pass filter exponent $\tau_p$ and the droop control parameter $k_p$ from grid-forming inverters: $H = \tau_p/k_p$, $\alpha=1/k_p$, $\forall i$ with $i=1,..,N$. 
The power fluctuation $\Delta P(t)$ is only applied at node $k$.
The impedance of the lines for typical medium-voltage (MV) grid lines with 20 kV base voltage equals $Z= Y^{-1}= (0.4 + 0.3j) \Omega/km$. The coupling strength between a node pair $(i,j)$ then equals $U_i|Y_{ij}|U_j$. 
The addition of line losses introduces a phase shift of $\phi_{ij}\approx \arctan(\frac{G_{ij}}{B_{ij}})$ that does not occur in the non-lossy model but will be shown to have significant consequences for stability. 

For each simulation run, the same intermittent time series was added to a single node's power input.  
The equation is written in the co-rotating frame, thus the synchronous state we study is characterized by $\omega_i = 0$. Hence, we are interested in the deviations from the stable frequency set point which corresponds to $50$Hz. In the following, we mainly investigate $\Delta f = \omega/(2\pi)$

%As I am interested in the low inertia case, with few low powered grid forming inverters at each node, I will assume a weakly reacting, strongly smoothed system. This leads me to consider $\alpha=0.01s$ and $H=0.1s^2$. Please note that the results are not sensitive to the exact choice of $\alpha$ and $H$.

An islanded power grid is balanced within itself, in our case there are 50 effective producers and  50 consumers with $P_i=\pm 0.2 MW$ power infeed before losses. Downstream each MV level node may lie a full Low-Voltage (LV) network which effectively every MV level node represents.  
%The power infeeds are chosen homogeneously to isolate topology and network effects in the model. As there is no connection to upper grid levels, losses are compensated locally at each node, and the net power infeed is given by $\tilde{P}_i = (P_i +P_{loss}/N)$.

\subsubsection{Intermittent Noise}
Important characteristics are the probability distribution function (PDF), the increment distribution and the power spectrum. 
If the PDFs, of both the time series of power and power increments, are fat tailed (the tails are  not exponentially bounded \cite{asmussen2008applied}), we define this as intermittency \cite{milan2013turbulent,anvari2016short}. Thus, time series from such sources show long-term temporal correlations. 
Also, the power generation from wind and solar power plants has a power spectrum that is power-lawed with the Kolmogorov exponent of turbulence.
The stochastic nature of such processes was identified in \cite{anvari2016short, schmietendorf2016} with the help of time series analysis.

In the following simulations the intermittent time series for solar and wind power fluctuations were generated by a clear sky index model, based on a combination of a Langevin and a Jump process, developed in \cite{anvari2016short}, and a Non-Markovian Langevin type model developed in \cite{schmietendorf2016}, respectively. Output of the models are wind and solar time series extrapolated to the resolution of the time-series analysis. Hence, we do not model stochastic differential equations explicitly. An example time series, $\Delta P(t)$, of the combined wind and solar power fluctuations is shown in Fig.\ref{fig:micro_net_exc} (left).
\subsection{Stability Measures}
We use the {\it exceedance} as our main stochastic stability measure to quantify the stability of the synchronous state. It is the cumulated time an observable stays outside a defined ``safe'' region \cite{feller1967introduction}. For our case we define a frequency threshold of $0.01$Hz. This threshold corresponds to the so-called dead band from the German transmission code which defines at which frequency primary control actions kick in to balance deviations from the desired $50$Hz set point \cite{transmission_code}.
As we apply single-node fluctuations for each run, $i=1,..,N$, and record the frequency response for each node $j=1,..,N$, we end up with $N\times N$ frequency time series from which stability measures are derived. One grid value represents the probability of network node $j$ to be outside the given frequency band when node $i$ is perturbed: 
\begin{equation}
E_{i,j}=P_i(|f_j|>0.01).
\label{eq:exc}
\end{equation}
This can be further aggregated into the nodal measures: 
\paragraph{Troublemaker Index ($\TI$)} The average exceedance over all $N$ nodes given a perturbation at $i$: 
	% Drivers of Mean Exceedance, dominance, troublemaker index, infectiousness
	\begin{equation}
	\TI=\bar{E}_i=\frac{1}{N}\sum_{j=1}^N E_{i,j}
	\label{eq:ti}
	\end{equation}
	Power fluctuations at a node with a high $\TI$ causes large frequency deviations often and/or at many nodes. 
	% , Exceedability, submissiveness
\paragraph{Excitability} quantifies how much a single-node is exceeding the frequency threshold on average when a random node in the network is perturbed:
	\begin{equation}
	\bar{E}_j=\frac{1}{N}\sum_{i=1}^N E_{i,j}
	\label{eq:excitability}
	\end{equation}
	Nodes with high excitability react strongly for many origins of the perturbation within the network. We call such nodes highly sensitive.

\subsection{Results}
The power flow losses on resistive lines induce an asymmetry in the coupling network. We find that this asymmetry induces several novel effects, including differentiating nodes that spread fluctuations throughout the network (trouble makers) and those that are excited by fluctuations no matter where they occur (excitable nodes). We derive predictors for the location of these nodes in the network.
\begin{figure*}[t!]
	\includegraphics[width=0.8\textwidth]{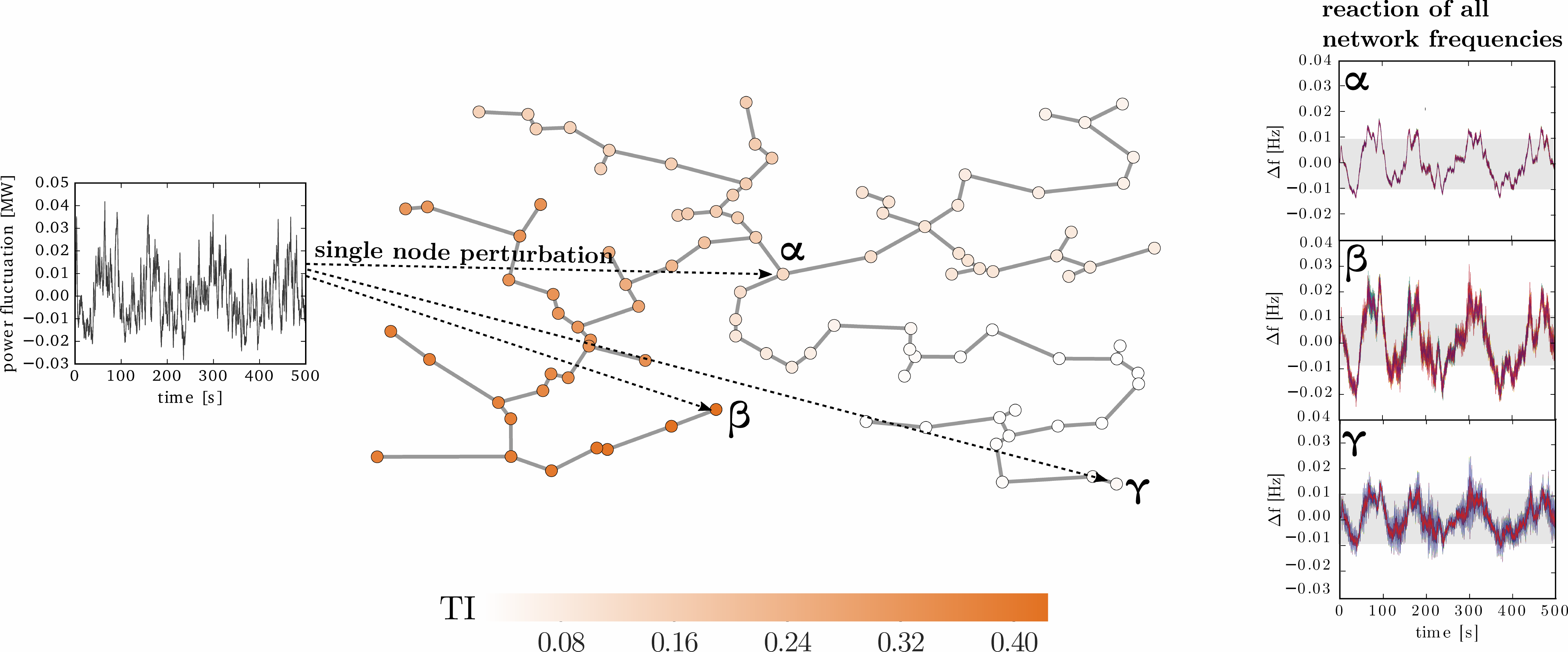}
	\caption{Left: Power fluctuation time series $\Delta P(t)$ jointly generated by solar and wind models that capture their intermittent behaviour \cite{anvari2016short,schmietendorf2016}. Center: Random microgrid with $\TI$ as colouring. One simulation run with single node fluctuations at one specific node produces this node's $\TI$ value  Right: Frequency time series for all network nodes $T=500$s for single-node fluctuations at node $\alpha$, $\beta$ and $\gamma$. The grey zone is the frequency threshold band of $0.01$Hz.}
	\label{fig:micro_net_exc}
\end{figure*}

\begin{figure}
	\centering
	\includegraphics[width=0.4\textwidth]{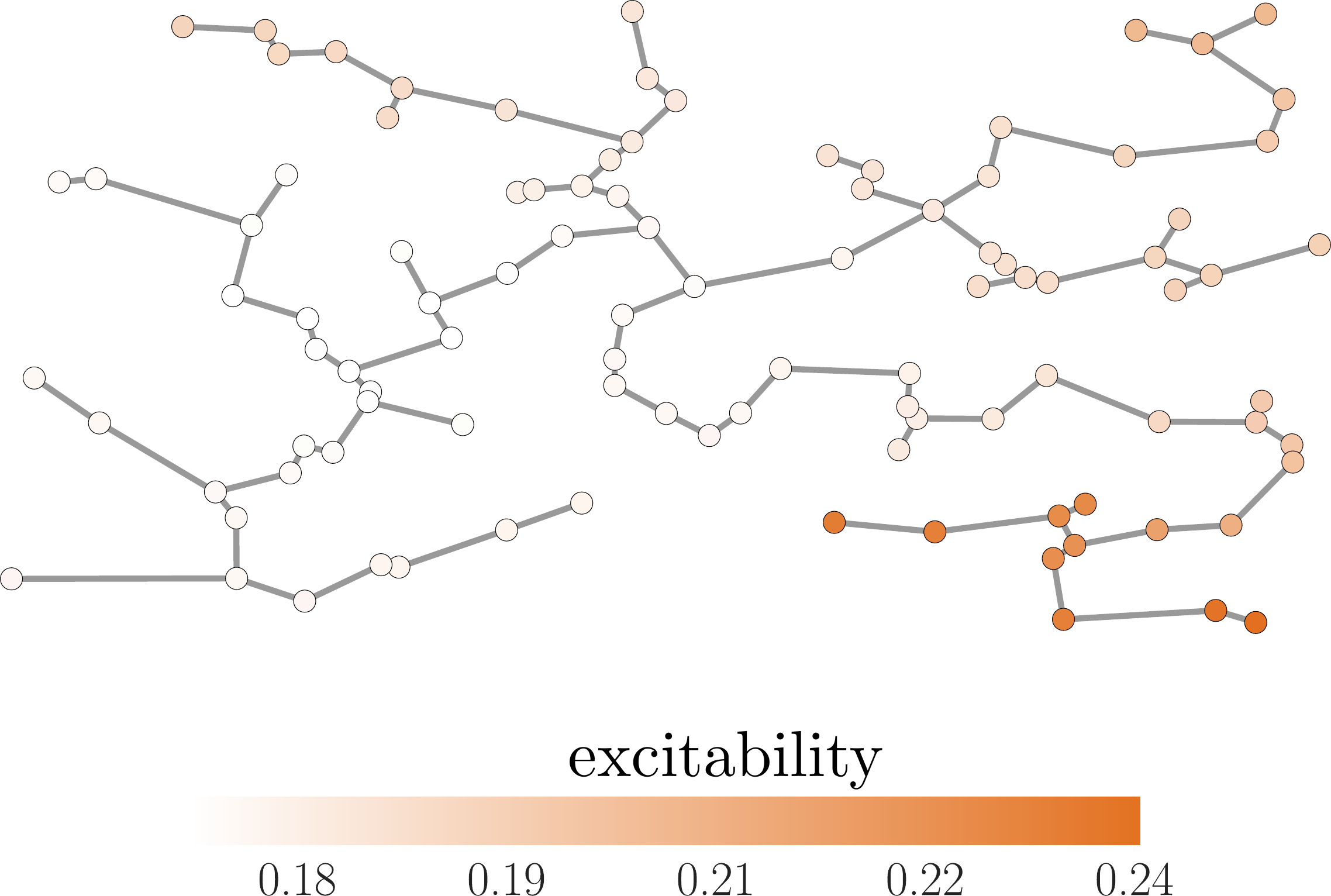}
	\caption{Excitability, see \eqref{eq:excitability}, for a synthetic islanded power grid.}
	\label{fig:micro_net_frq_disp}
\end{figure}

Figs. \ref{fig:micro_net_exc} and \ref{fig:micro_net_frq_disp} shows how the the power grid splits into branches of troublemaker nodes and of fluctuation sensitive nodes. Both types of nodes can be identified with the eigenvectors of the network's Jacobian $\mathbf{J}$. 
\begin{align}
\mathbf{J}&=
\begin{bmatrix}
\frac{\partial \dot{\phi_i}}{\partial\phi_j}&\frac{\partial \dot{\phi_i}}{\partial\omega_j}\\
\frac{\partial \dot{\omega_i}}{\partial\phi_j}&
\frac{\partial \dot{\omega_i}}{\partial\omega_j}
\end{bmatrix}=\begin{bmatrix}
\mathbf{0}_{N\times N} & \mathbf{1}_{N\times N}\\
\frac{1}{H} \mathbf{L} & -\frac{\alpha}{H} \mathbf{1}_{N\times N}
\end{bmatrix},
\label{eq:jac}
\end{align}
Since losses eliminate the symmetry of the weighted Laplacian matrix, $\mathbf{L}$, in the model system, it is necessary to distinguish between left and right eigenvectors for the non-symmetric Laplacian. These appear in orthonormal pairs with the same corresponding eigenvalue: %Thus, for vector $\sum_j a^jv_r^{kj}$ wird auf sich selbst abgebildet.}
\begin{align*}
v^i_r\cdot v^j_l &= \delta^{ij},\\
L^{km}&=\sum_i v^{ki}_r \lambda ^i v^{mi}_l.
\end{align*}

\paragraph{Left eigenvectors and troublemakers.} \label{subsec:excitablenodes_predictor}
In \cite{auer2018stability} it was shown that the strongly autocorrelated drivers of instability or troublemakers are most visible in the zero mode. Hence, the predictor for the troublemaker index, $Pred_{\TI}$, is proportional to the left eigenvector of the zero eigenmode:
\begin{equation}
Pred_{\TI}= v_l^0\;.
\label{eq:pred_TI}
\end{equation}    
Heuristically, this can be understood in the following way: In order to get a large frequency deviation, it requires a sustained displacement of the system. The zero eigenmode is the only one that does not lead to oscillations of the system which cancel each other out. This holds if the system is not overdamped. Thus, there is only a chance of a sustained perturbations for the zero mode. This predictor is derived from a sound analytic calculation with a linear response approach and  shows an excellent correlation with the troublemaker index from the simulation results \cite{auerplietzsch2018}.

\paragraph{Right eigenvectors and excitable nodes.} \label{subsec:excitablenodes_predictor}
Conversely, in \cite {auer2018stability} it was shown that the right excitability is given by the %that temporal correlations play no such prominent role when considering excitability. Instead, the spatial structure of the modes becomes relevant, which is encoded in the 
right eigenvectors, $v^i_r$, of the Laplacian.
The predictor for excitability, derived from analytic calculations \cite{auerplietzsch2018}, contains the sum over all right eigenvector modes since it gives insights on how strong each node is excited on average. 
Because intermittent RES fluctuations show a turbulent power spectrum, with the Kolmogorov exponent of $-5/3$,  not all network modes are excited equally strong and the right eigenvectors needs to be weighted. Hence, intermittent noise especially excites the lower harmonics of the networks' eigenmodes. Then, the excitability predictor $\text{Pr}_{\bar{E}_m}$ of node $m$ is defined as

\begin{align}
\begin{split}
\text{Pr}_{\bar{E}_m} &= \sum_i |v_r^{m,i}| n(\lambda_i)\\
\end{split}
\label{eq:pred_exc}
\end{align}
where 
\begin{align}
\begin{split}
n(\lambda)=\left\{
\begin{array}{ll}
|\lambda|^{-5/3} \ &for\ \lambda\neq0,\\
0 \ &for \ \lambda=0.
\end{array}
\right.
\end{split}
\end{align}

Again, this predictor correlates well with the simulation results, as shown in \cite{auer2018stability}.

\section{Delayed Control and Transient Stablity}
\label{sec:delay}
This section presents research on the impact of delayed reaction on power grid stability, which relates to the question of how to enable a dynamically stable integration of renewable energies.

In the future both consumption and production may be controlled by power electronic devices. These devices, such as grid-following inverters, need measurement and reaction times and hence, introduce delays into the power grid dynamics \cite{naduvathuparambil2002communication}. Delays of the order of several hundred milliseconds might arise in addition to delays of unknown magnitude caused by demand response.
ENTSO-E acknowledges the potential challenges arising from delayed control of power electronic devices in their 2016 guidance document for national implementation for network codes on grid connection \cite{HPoPEIPs}.

Hence, the following investigation considers a large range of potential delays $\tau \in (0,5)$s looking for the boundary of acceptable delays. The transient stability of the power grid frequency is evaluated with linear stability analysis the study of Lyapunov exponents \cite{pikovsky2003synchronization}) and time-domain analysis with Monte Carlo simulations (basin stability). 
The Basin Stability (BS) of a multi-stable dynamical system quantifies the trajectories that approach the synchronous state after random initial perturbations at single nodes \cite{Menck2013}. Unstable paths  will lead the system to a different, undesirable attractor.
Much of this section is based on \cite{auer2018stability,Schaefer2015,schafer2016taming}. \footnote{\textcopyright~EDP Sciences and Springer 2016. With permission of Springer.}.

\subsection{Model Description}\label{sec:dsgc_model_descr}
In order to be able to undertake both analytic calculations and simulations for a four-node star network with  one larger producer (with $P=3 pu$) and three consumers, $P_i=1 pu$, was chosen. Larger networks up to 9 nodes were investigated as well with the same effects of instability occurring. With larger networks computation time becomes a limiting factor.

The equation of motion for power system dynamics with delayed control is derived by introducing a power adaption term that follows frequency deviations with a time lag $\tau$:
\begin{align}
\begin{split}
\label{eq.motion with delay}
\frac{\mathrm{d}^2\phi_i}{\mathrm{d}t^2}&=P_i-\alpha_i\frac{\mathrm{d}\phi_i}{\mathrm{d}t}+\sum_{j=1}^N K_{ij}\sin(\phi_j-\phi_i)
\\&-\gamma_i\frac{\mathrm{d}\phi_i}{\mathrm{d}t}(t-\tau),
\quad \forall i \in \{1,...,N\}.
\end{split}
\end{align}
For the parameteriziation of the model case please read \cite{Schaefer2015}.
%The moment of inertia is assumed to be identical for all machines \cite{Schaefer2015}. As a further simplification, ohmic loads are neglected as they should be small compared to shunt admittances \cite{VanHertem2006} for the second-scale dynamics. The parameters of the swing equation are calculated from standard literature values \cite{Machowski2011,schultz2014detours}.
%Delayed differential equations need a history function as initial condition that describes the system dynamics for the time interval $[-\tau,0]$:
%\begin{equation}
%\omega(t<0)=\omega_0 (1+0.1\tanh(t/2)).
%\label{eq:history_func}
%\end{equation}
%The simulation results do not change significantly with different history functions such as $\omega(t<0)=\omega_0$. Though, the choice of \eqref{eq:history_func} has the advantage to be smooth, non-constant and thus more realistic.

As an addition to the model, the potentially stabilizing effect of signal averaging was taking into account by averaging frequency measurements over time intervals of lengths $T$. Such averaging yields:
\begin{align}
\label{eq.motion with average}
\frac{\mathrm{d}^2\theta_i}{\mathrm{d}t^2} & = P_i-\alpha_i\frac{\mathrm{d}\theta_i}{\mathrm{d}t}+\sum_{j=1}^N K_{ij}\sin(\theta_j-\theta_i)\\
&-\frac{\gamma_i}{T}\int_{t-T}^{t}\frac{\mathrm{d}\theta_i}{\mathrm{d}t}(t'-\tau) \mathrm{d}t'.\\
\end{align}
To simplify the analytic calculations, averaging times $T$, delays $\tau$ as well as $\alpha$ and $\gamma$ were chosen homogeneously for all nodes. In addition, line coupling or capacities $K_{ij}=K$ are chosen homogeneous for all lines.  
In the following sections the delayed equations \eqref{eq.motion with delay} and \eqref{eq.motion with average} are evaluated according to their stability as a function of the delay $\tau$ with different averaging times $T$.

\subsection{Impact of Delays on Power Grid Stability}
\begin{figure}[t!]
	\centering
	\includegraphics[width=0.47\textwidth]{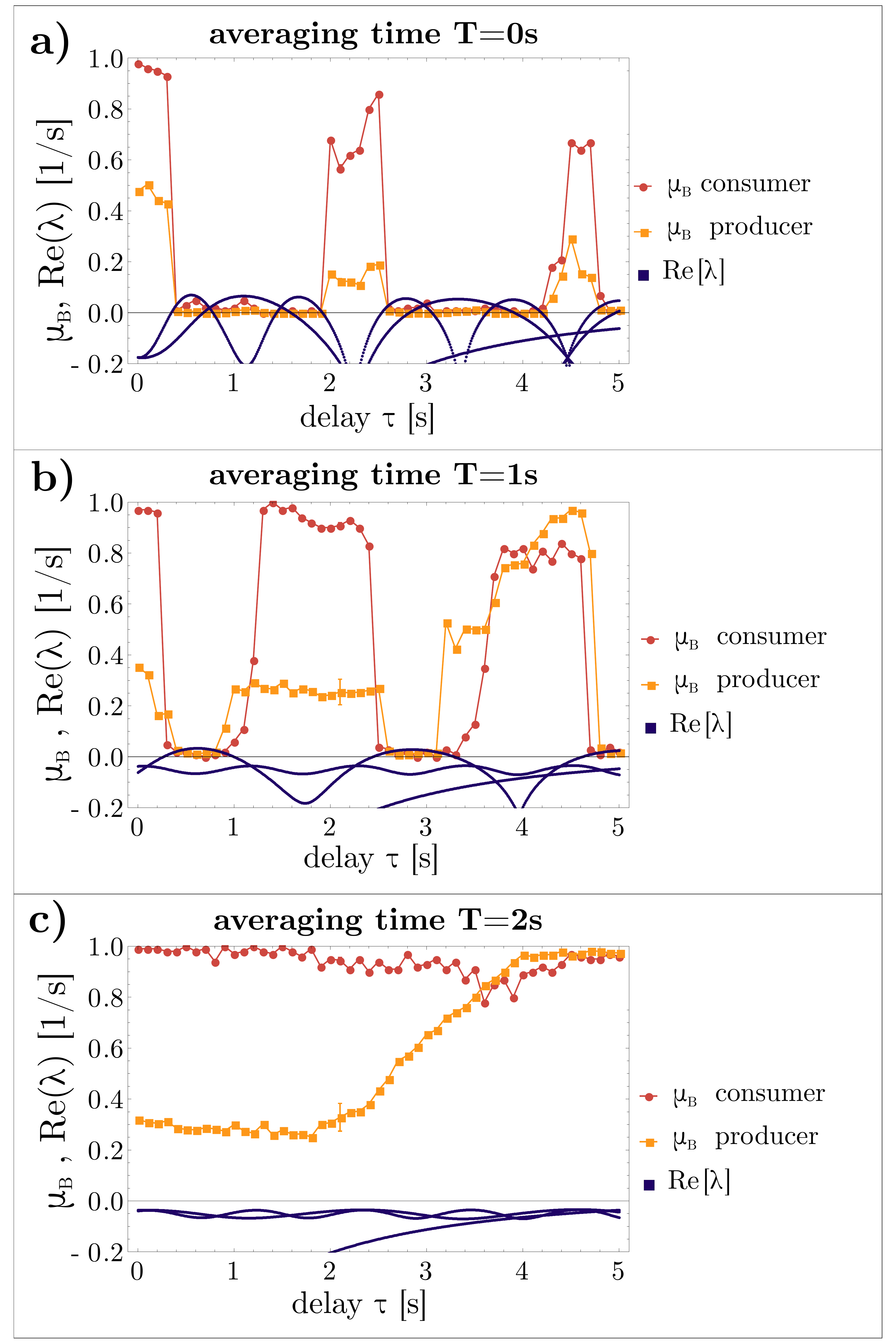}
	\caption[Linear stability and basin stability analysis for the star topology]{\label{fig:4_node_star_linear_and_basin_Stability}
		\textbf{Linear stability and basin stability analysis for the star topology}. Intermediate delays result in high values of basin stability if averaging is switched on.
		Shown are the real parts of the eigenvalues for the 4 node star motif (dark blue) as well as the basin stability of the producer (orange) and of one consumer (red) as functions of the delay $\tau$ for different averaging times: \textbf{a)} $T=0\text{s}$, \textbf{b)} $T=1$s, \textbf{c)} $T=2$s. Parameters $\alpha =0.1/\text{s}$, $K=8/\text{s}^2$ and $\gamma =0.25/\text{s}$ were applied. For delay $\tau=2.1$s simulations were repeated 21 times, averaged and the standard deviation is shown as a typical error bar. Figures are taken from \cite{Schaefer2015}.}
\end{figure}

In \cite{Schaefer2015}, it was shown that a delayed system poses risks to the stability of the power grid for certain delays $\tau$. Fig. \ref{fig:4_node_star_linear_and_basin_Stability} shows the stability impact of delays for a four-node star motif and compares the results of linear stability and basin stability analysis.  The authors have found destabilizing effects of resonances and the "rebound effect" for large delays and presented how intermediate delays $\tau$ benefit the stability. 
Hence, the stability properties depend crucially on the delay and the averaging applied.
Without any averaging there are delays $\tau$ for which the fixed point is linearly unstable, i.e., there are eigenvalues with a positive real part $\Re({\lambda}) \ge 0$ (see the blue curve of Fig. \ref{fig:4_node_star_linear_and_basin_Stability}a). Those eigenvalues exhibit a periodic behavior with respect to the delay $\tau$. Operating the power grid at a delay $\tau$, e.g. $\tau \approx 1$s for which the real part of the eigenvalue is positive, is equivalent to resonantly driving the power grid away from the fixed point instead of damping it towards stable operation. These destabilizing delays are linked to the eigenfrequency of the oscillators in the power grid. If the delay is half the eigenoscillation duration, then it increases amplitudes of perturbations instead of damping them. This destabilization only occurs for $\alpha<\gamma$ because the resonant driving has to be larger than the intrinsic damping of the system, see also \cite{Schaefer2015}.

With an extension using frequency measurements averaged over time intervals of lengths $T$ the power grid could be stabilized regardless of the specific delay. 
Introducing sufficiently large averaging times into the control cures these instabilities (Fig. \ref{fig:4_node_star_linear_and_basin_Stability}b and c).

\section{Open-source Dynamic Power System Modeling}
\label{sec:os-modeling}
The two research subjects described above shall illustrate limitations and hurdles that need to be overcome in state-of-the-art open-source dynamic modeling of power grids with high shares of renewable energies. These limitations refers to both the programming language, Python, but also to the unavailability of open-source libraries for modeling power grid dynamics. In the following, the advantages of Julia compared to other open-source languages will be sketched in short. Then, the next subsection presents the OS library PowerDynamics.jl in short with its current implementation status and the envisioned functionalities. For a more detailed introduction into PowerDynamics.jl the interested reader may be referred to \cite{tkittelWIW2018}.

\subsection{Why Julia?}
\paragraph*{Solvers} Julia has a rich set of solvers that represent an important prerequisite for dynamic power grid modeling. For renewable generator dynamics Julia is able to capture the stochasticity of intermittent power fluctuations with Stochastic Differential Equations. Even highly flickering photovoltaic power production can be modeled as Jump Diffusion equations (stochastic differential equations with discontinuous). So far, an easy integration of such processes was not possible in Python.

The same holds true for Delay Differential Equations (DDEs) to model delayed reaction from power electronic devices. In Python it would be necessary to integrate not well maintained c-coded libraries into modeling projects. The alternative is to use commercial software such as Modelica or Matlab to solve such problems. In Julia, with DelayDiffEq.jl finally a working open-source implementation is available.

Also, the integration of loads as algebraic constraints with Differential Algebraic Equations in Julia can be easily solved with DASSL.jl. This allows to preserve the network structure without using Kron-reduced power grid models \cite{dorfler2013kron}.

\paragraph{Performance} Using the just-in-time (JIT) compilation method, Julia produces efficient native code for multiple platforms at runtime.
\paragraph{Rapid software development} It uses dynamic typing, so it is easy to use, feels like a scripting language and has good interactive use without losing any performance. That way, it saves the developer a lot of time.
\paragraph{Scientific computing} Julia has been developed with scientific computing in mind. It's syntax focuses on precise mathematics. Many datatypes and even parallelism is available out of the box.
\paragraph{Generality} Using multiple dispatch, Julia allows object-oriented and functional programming patterns at the same time.
\paragraph{Julia is composable} Julia has been designed such that independent packages work well together without any extra work.
% TIM: I took the following out as it is basically your first point.
%\paragraph{Julia has a growing ecosystem} The aforementioned advantages induces a growing community and with that a growing ecosystem of Julia packages. Particularly important for modeling is the \diffeq{} library which provides high-performance solvers and interfaces to industrial grade solvers like \sundials{} \cite{hindmarsh2005sundials} for solving multiple kinds of differential equations (ordinary, algebraic, delayed, stochastic).
\paragraph{Metaprogramming} At execution time, the source code can be modified automatically and one can even add own syntax to Julia. 

\subsection{Why PowerDynamics.jl?}
In PowerDynamics.jl we want to make use of all the above described advantages of the Julia programming language. So far, the general framework of the software library is set up \cite{tkittelWIW2018} with a few already implemented bus types. This included for examples the 2nd- and 4th-order Synchronous machine model and ohmic loads as algebraic constraints \cite{Sauer2006,Kundur1994}.
We plan to further implement several different inverter control types, detailed dynamic line models and DDE nodes. These different components shall then be practically validated via test bed simulations and measurements. Here, the goal is to take the newest regulatory standards into account \cite{TAR4110,TAR4120}.

The advantage of an open-source library for such simulations is the transparency and thus credibility of the underlying model framework. Of course, this does not imply a necessary publication of the network data which is sensible data.

\section{Summary \& Conclusion} \label{sec:discussion}

\subsection{Intermittent Renewables in PowerDynamics.jl}
In \cite{auer2017stability, auer2018stability,auerplietzsch2018} we have used synthetic power grids to understand the qualitative behavior of frequency dynamics with intermittent renewable power infeed. We have found that the combination of resistive lines leads to a strongly asymmetric frequency deviations across the network nodes, depending on where fluctuations were initially introduced. Further, the intermittent nature of renewable power fluctuations leads to an excitation of the lower harmonics of the network eigenmodes. Analytically, we have found predictors for the nodes causing relatively strong frequency deviations across all network nodes (troublemaker nodes) and for nodes that are relatively easy excitable irrespective of where perturbations were initially introduced. For simplicity we used single node perturbations. However, the analytic calculations were undertaken with linear response theory and thus, the dynamics from multiple node fluctuations can be easily integrated by the superposition of the single-node results.

In PowerDynamics.jl we also want to model the influence of renewable power fluctuations on power quality and extend the insights from the result described above. The computational performance will allow the implementation of real-world power grids with frequency dynamics given by different inverter control schemes and power fluctuations explicitly modeled with SDEs and Jump Diffusion processes. 

\subsection{Delayed Control in PowerDynamics.jl}
In \cite{Schaefer2015} and \cite{schafer2016taming} we have shown the potential instabilities caused by delayed reaction from power electronics in the power grid. These are caused by the excitation of different network eigenmodes that appear periodically with changing delay. The results are robust for different network motifs \cite{schafer2016taming}. However, due to computational constraints we were restricted to simple four-node start motifs, small ring or lattice structures.

PowerDynamics.jl will allow to model DDEs with larger node number and hence the usage of model cases from IEEE test grids. This will be of great importance to test the relevant time scales for larger grids and not only network motifs with different network eigenmodes and thus resonance frequencies.
\subsection{Further Questions to be answered with PowerDynamics.jl}
There is variety of further interesting questions to be answered with our new modeling tools in hands. This includes the question of the necessary model detail for reliable stability analysis, especially for distribution grids where there is a high number of consumers, producers or even producers and a lack of information about dynamic grid components at the same time.
Also, the question about the necessary share of virtual inertia in islanded power systems (and also future continental grids) is still an open issue.
To name only one more example, the interaction of different inverter control schemes is up to now not fully understood and needs to be modeled in great detail.

All in all, transitioning to PowerDynamics.jl will allow us to move from academic research questions with modeling qualitative results to a sort of on-road test for the relevance of such results in today's renewable distribution grids.

\section*{Acknowledgment}

This paper was presented at the 19th Wind Integration Workshop and published in the workshop’s proceedings.

We would like to thank the German Academic Exchange Service for the opportunity to participate at the Wind Integration Workshop 2018 in Stockholm via the funding program ``Kongressreisen 2018''.
Further, the authors are currently funded by the Climate-KIC Pathfinder project ``elena -- electricity network analysis'' by the European Institute of Innovation \& Technology.
This work has been conducted within the Complex Energy Networks research group at the Potsdam Institute for Climate Impact Research.

We would like to thank Frank Hellmann and Paul Schultz for the discussions on structuring an Open-Source library for dynamic power grid modeling.



%\bibliographystyle{IEEEtran}

\bibliographystyle{IEEEtran}
\bibliography{Literature_joint.bib}

\end{document}